\documentclass[aps,reprint,10pt,notitlepage,pra,
               onecolumn,
               amsfonts,amssymb,amsmath,
               showpacs,
               nobibnotes,
               nofootinbib,
               ]{revtex4-1}
\usepackage{graphicx}
\usepackage{amssymb}
\usepackage{amstext}
\usepackage{amsmath}
\usepackage{amsthm}
\usepackage[warn]{mathtext}
\usepackage{braket}
\usepackage{bm}
\usepackage{indentfirst}
\usepackage{wrapfig}
\usepackage{topcapt}
\usepackage{comment}
\usepackage{subfigure}
\usepackage{color,xcolor}
\definecolor{dark-blue}{rgb}{0,0,0.875}
\definecolor{dark-green}{rgb}{0,0.625,0}
\definecolor{dark-red}{rgb}{0.875,0,0}
\usepackage[colorlinks=true,linkcolor=dark-red,filecolor=dark-blue,
citecolor=dark-green,urlcolor=dark-blue,breaklinks=true]{hyperref}
\begin{document}
\title{Generation of entangled states of light using discrete solitons in waveguide arrays}
\author{V. O. Martynov}
\affiliation{Institute of Applied Physics of the Russian Academy of Sciences, Nizhny Novgorod, 603950, Russia}
\author{V. O. Munyaev}
\affiliation{Department of Control Theory, Nizhny Novgorod State University,
	Gagarin Av. 23, 606950, Nizhny Novgorod, Russia}
\author{L. A. Smirnov}
\affiliation{Institute of Applied Physics of the Russian Academy of Sciences, Nizhny Novgorod, 603950, Russia}
\affiliation{Department of Control Theory, Nizhny Novgorod State University,
Gagarin Av. 23, 606950, Nizhny Novgorod, Russia}

\begin{abstract}
We study the quantum properties of light propagating through an array of coupled nonlinear waveguides and forming a discrete soliton. 
We demonstrate that it is possible to use certain types of quasi-solitons to form continuous variables entanglement between the certain pair of waveguides.
This entanglement generation process occurs even with a very large number of photons in the laser field used, so it does not require a special material with an extremely high nonlinearity coefficient. 
Also absorption in the waveguide media does not influence it too much.
\end{abstract}
\maketitle
On-chip integrated optical waveguide circuits open a wide variety of possibilities to manipulate quantum states of light~\cite{zoubi_quantum_2017, silverstone_-chip_2014, matthews_manipulation_2009, kruse_dual-path_2015}. 
One of the most promising usage of such devices is implementation of various quantum informatics algorithms~\cite{nielsen_quantum_2000}. Entangled states of light are the key element of quantum information processing. 
Development of integrated devices which generate such states using ordinary laser sources as their input appears to be an important problem. 
Recently, there have been proposed a number of schemes~\cite{guo_parametric_2017, caspani_integrated_2017, solntsev_path-entangled_2017, yang_manipulation_2014} to form entanglement between optical modes in different waveguides. 
These schemes are based on down-conversion processes in nonlinear media. 
They require the use of specific materials and waveguide geometry, which complicates the design of the discussed devices. 
An alternative approach might be to use nonlinear self-action process which is more common and occurs in all materials with cubic nonlinearity. 
Currently, the possibility of using this process to generate entangled states is poorly studied. 
Here we propose to use an array of coupled waveguides with cubic nonlinearity to generate continuous variables entanglement~\cite{weedbrook_gaussian_2012}. 
Such objects are well studied from classical nonlinear dynamics point of view. 
It is known that there is a large variety of discrete solitons in such arrays~\cite{kevrekidis_discrete_2009, christodoulides_discretizing_2003}.
Taking quantum effects into account, the propagation of such field distributions can lead to nonclassical photon statistics. 
Such effect is known for continuous solitons~\cite{corney_quantum_2001, lai_entangled_2009}.
We will show that the similar behavior present in the discussed system.
The basic idea is that due to nonlinearity, the photon statistics will change as light travels through the waveguide.
And the interaction between the optical modes of the waveguides will transform this non-classical state of light into an entangled one. 
\par
\begin{figure}[h!]
	\centering\includegraphics[width=7cm]{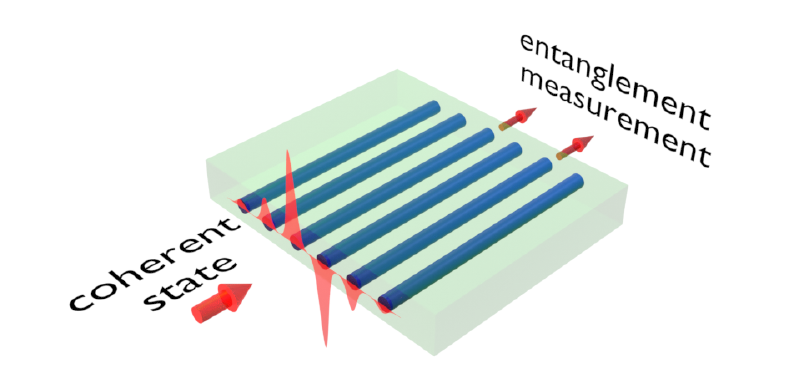}
	\caption{Schematic of quantum light propagation along an array of nonlinear waveguides.}
	\label{fig:schema}
\end{figure}

We consider an array consisting of $N$ single-mode waveguides~(Fig.~\ref{fig:schema}). 
A long laser pulse is sent to the input of each waveguide. Propagation of the pulse may be described in terms of slowly varying field operators $\hat{\psi}_{k}\left(x,t\right)$, where $k$ is the index of a waveguide, $x$ is the coordinate along array. The Hamiltonian of system consists of two parts:
\begin{equation}\label{eg:base_ham}
	\hat{H}=\hat{H}_s + \hat{H}_i.
\end{equation}
$\hat{H}_s$ refers to the independent propagation of light along individual waveguides.

In the slowly varying envelope and rotating-
wave approximations it equals~\cite{drummond_quantum_2014}: 
\begin{equation}
	\hat{H}_s = \sum^{N}_{k=1}\int dx\left[\dfrac{iv}{2}\left(\hat{\psi}^{\dag}_k\nabla\hat{\psi}^{\phantom{\dag}}_k - \nabla\hat{\psi}^{\dag}_n\hat{\psi}^{\phantom{\dag}}_k\right) +\dfrac{U}{2}\hat{\psi}^{\dag2}_k\hat{\psi}^{\phantom{\dag}2}_k\right], 
\end{equation}
v is group velocity. Hereinafter we assume that $\hbar=1$. $\hat{H}_i$ is the Hamiltonian which describes interaction of optical field in different waveguides. In case of weak coupling it equals~\cite{agarwal_quantum_2012}:
\begin{equation}
	\hat{H}_i = \sum^{N-1}_{k=1}\int \kappa\hat{\psi}^{\dag}_k\hat{\psi}^{\phantom{\dag}}_{k+1} dx +\rm{H.c.}
\end{equation}
Using Hamiltonian~\eqref{eg:base_ham} one can derive Heisenberg equations:
\begin{equation}
	\begin{array}{r}
		\left[-v\dfrac{\partial}{\partial x} + \dfrac{\partial}{\partial t}\right]\hat{\psi}^{\phantom{\dag}}_k\!\left(x,\!t\!\right) = -i\kappa\left[\hat{\psi}^{\phantom{\dag}}_{k-1}\!\left(x,\!t\!\right) + \hat{\psi}^{\phantom{\dag}}_{k+1}\!\left(x,\!t\!\right)\right] \\ -iU\hat{\psi}^{\dag}_k\!\left(x,\!t\!\right)\hat{\psi}^{\phantom{\dag}}_{k}\!\left(x,\!t\!\right)\hat{\psi}^{\phantom{\dag}}_{k}\!\left(x,\!t\!\right).
	\end{array}
\end{equation}
To simplify further calculations, we perform them in a moving frame using substitution $t_v = t - x/v$. Also we normalize distance $x \rightarrow z = \kappa x / v$:
\begin{equation}\label{eq:heis}
	\begin{array}{r}
		\dfrac{\partial}{\partial z}\hat{\psi}^{v}_k\!\left(z,\!t_v\!\right) = i\left[\hat{\psi}^{v}_{k-1}\!\left(z,\!t_v\!\right) + \hat{\psi}^{v}_{k+1}\!\left(z,\!t_v\!\right)\right] \\ +iL\hat{\psi}^{\dag v}_k\!\left(z,\!t_v\!\right)\hat{\psi}^{v}_{k}\!\left(z,\!t_v\!\right)\hat{\psi}^{v}_{k}\!\left(z,\!t_v\!\right),
	\end{array}
\end{equation}
where $L=Uv/\kappa$ and
\begin{equation}
	\hat{\psi}^{v}_k\!\left(z,\!t_v\!\right) = \hat{\psi}^{\phantom{\dag}}_k\!\left(\dfrac{vz}{\kappa},\!t - \dfrac{z}{\kappa}\!\right).
\end{equation}
$t_v$ in equations~\eqref{eq:heis} $t_v$ s just a parameter so in further calculations we omit it assuming that it is always the same and corresponds to the center of the input pulses. 
To calculate evolution of light field along the array based on equations~\eqref{eq:heis} we assume that quantum state of light is Gaussian for every value of $v$. 
This mean, that Sudarshan-Glauber P distribution has Gaussian profile. 
For such distributions cumulants~\cite{gardiner_stochastic_2009} $\ll\hat{X}_1\cdot\hat{X}_2\cdot\hat{X}_3\gg$ and $\ll\hat{X}_1\cdot\hat{X}_2\cdot\hat{X}_3\cdot\hat{X}_4\gg$ equal to zero if $\hat{X}_l$ takes a value from the set $\left\{\hat{\psi}^{v}_{k}, \hat{\psi}^{\dag v}_{k} \right\}$ in such a way that the products are normally ordered. This fact allows using the following relations:

\begin{equation}
	\begin{array}{r}
		\langle \hat{X}_1\!\cdot\!\hat{X}_2\!\cdot\!\hat{X}_3\rangle = \langle\hat{X}_1\rangle\langle\hat{X}_2\!\cdot\!\hat{X}_3\rangle + \langle\hat{X}_2\rangle\langle\hat{X}_1\!\cdot\!\hat{X}_3\rangle  \,\,\,\,\,\,\,\,\,\,\,\,\,\,\,\,\,\,\,\,\,\,\,\,\,\\+ \langle\hat{X}_3\rangle\langle\hat{X}_1\!\cdot\!\hat{X}_2\rangle - 2\langle\hat{X}_1\rangle \langle\hat{X}_2\rangle\langle\hat{X}_3\rangle \\
		\langle \hat{X}_1\!\cdot\!\hat{X}_2\!\cdot\!\hat{X}_3\!\cdot\!\hat{X}_4\rangle \!=\! \langle\hat{X}_1\!\cdot\!\hat{X}_2\rangle\langle\hat{X}_3\!\cdot\!\hat{X}_4\rangle \!+\!  \langle\hat{X}_1\!\cdot\!\hat{X}_3\rangle\langle\hat{X}_2\!\cdot\!\hat{X}_4\rangle \\+ \langle\hat{X}_1\!\cdot\!\hat{X}_4\rangle\langle\hat{X}_2\!\cdot\!\hat{X}_3\rangle - 2\langle\hat{X}_1\rangle \langle\hat{X}_2\rangle\langle\hat{X}_3\rangle\langle\hat{X}_4\rangle
	\end{array}
\end{equation}
to derive a closed set of equations for quantities $\alpha_k = \sqrt{L} \langle\hat{\psi}^{v}_{k}\rangle$, $\Delta^{a}_{k,l} = L \langle\hat{\psi}^{v}_{k}\hat{\psi}^{v}_{l}\rangle - \alpha^{\phantom{*}}_k\alpha^{\phantom{*}}_l$ and $\Delta^{n}_{k,l} = L \langle\hat{\psi}^{\dag v}_{k}\hat{\psi}^{v}_{l}\rangle - \alpha^{*}_k\alpha^{\phantom{*}}_l$~\cite{tikhonenkov_quantum_2007}: 
\begin{equation}\label{eq:second_order}
	\begin{array}{c}
		\dfrac{\partial}{\partial z}\alpha_{k}\! =\! i\left(\alpha_{k-1}\! +\! \alpha_{k+1} \right) \!+\! i\left|\alpha^{\phantom{*}}_{k}\!\right|^{2}\alpha^{\phantom{*}}_{k} \!\!+\! i\left(2\alpha^{\phantom{*}}_{k}\Delta^{n}_{k,k} \!+\! \alpha^{*}_{k}\Delta^{a}_{k,k}\right),\\
		\dfrac{\partial}{\partial z}\Delta^{n}_{k,l} = i\left(\Delta^{n}_{k,l+1} + \Delta^{n}_{k,l-1} - \Delta^{n}_{k+1,l} - \Delta^{n}_{k-1,l}\right) \\+ 2i\left(\Delta^{n}_{l,l} - \Delta^{n}_{k,k} + \left|\alpha^{\phantom{*}}_{l}\!\right|^{2} -\left|\alpha^{\phantom{*}}_{k}\right|^{2}\right)\Delta^{n}_{k,l}  \\
		+ i\left(\Delta^{a*}_{k,l}\Delta^{a}_{l,l} - \Delta^{a}_{k,l}\Delta^{a*}_{k,k} + \alpha^{2}_{l}\Delta^{a*}_{k,l} - \alpha^{*2}_{k}\Delta^{a}_{k.l}\right), \\
		\dfrac{\partial}{\partial z}\Delta^{a}_{k,l} = i\left(\Delta^{a}_{k,l+1} + \Delta^{a}_{k,l-1} + \Delta^{a}_{k+1,l} + \Delta^{a}_{k-1,l}\right)  \\+ 2i\left(\Delta^{n}_{k,k} + \Delta^{n}_{l,l} + \left|\alpha^{\phantom{*}}_{k}\right|^{2} + \left|\alpha^{\phantom{*}}_{l}\!\right|^{2}\right)\Delta^{a}_{k,l}  \\
		+ i\left(\Delta^{a}_{k,k} + \alpha^{2}_{k}\right)\Delta^{n}_{k,l} + i\left(\Delta^{a}_{l,l} + \alpha^{2}_{l}\right)\Delta^{n*}_{k,l} \\+ i\dfrac{L}{2}\delta_{k,l}\left(\alpha^{2}_{k} + \alpha^{2}_{l} + \Delta^{a}_{k,k} + \Delta^{a}_{l,l}\right).
	\end{array}
\end{equation}
\par
\par
\begin{figure}[h!]
	\centering\includegraphics{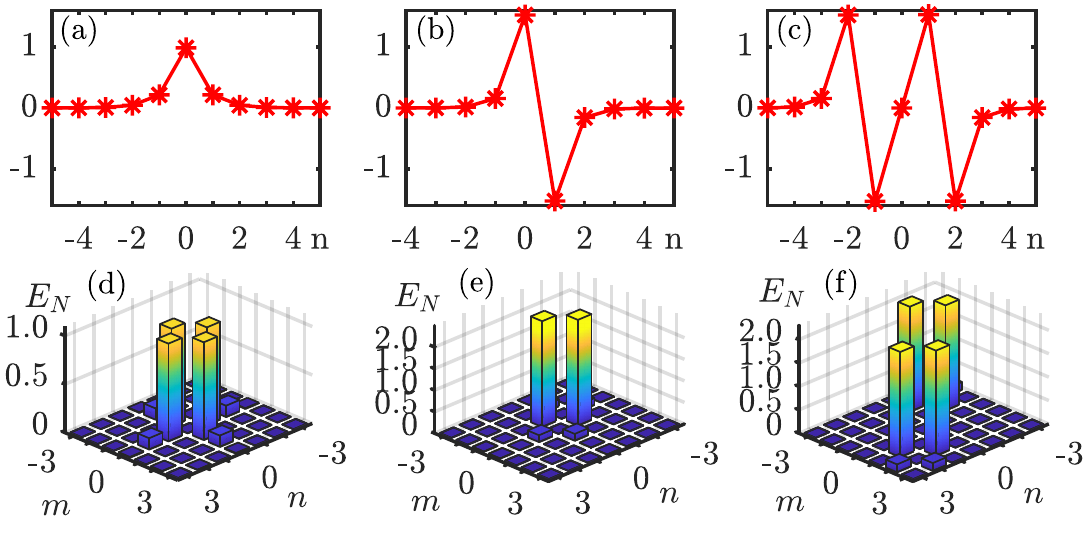}
	\caption{Top row represents the structure of discrete soliton solutions for light propagation in an array of waveguides with cubic nonlinearity. The bottom row shows the distributions of bipartite entanglement between different pairs of waveguides for the corresponding discrete soliton from the top row. The logarithmic negativity~($E_N$) is used to quantitatively describe entanglement. }
	\label{fig:structure}
\end{figure}
Before we start solving the equations~\eqref{eq:second_order}
we need to define boundary conditions at plane $z=0$. 
We assume that the quantum state of light is a multi mode coherent state at the input of the array. For such sort of states $\Delta^{a}_{k,l}$ and $\Delta^{n}_{k,l}$ equal zero for every $k,l = 1\ldots N$. So it is enough to specify the distribution of values $\alpha_k\!\left(z=0\right)$. Before we do so it should be noted that if in~\eqref{eq:second_order} one set $L=0$ the system of equations will be reduced to:
\begin{equation}\label{eq:classic}
	\dfrac{\partial}{\partial z}\alpha_{k}\! =\! i\left(\alpha_{k-1}\! +\! \alpha_{k+1} \right) \!+\! i\left|\alpha^{\phantom{*}}_{k}\!\right|^{2}\alpha^{\phantom{*}}_{k}.
\end{equation}
We want to emphasize here that $ L = 0 $ does not mean that we are neglecting nonlinearity. 
Instead, it means that the number of photons in optical modes tends to infinity, which can be interpreted as the transition to the classical physics limit. 
Equations~\eqref{eq:classic} have solutions which are called discrete solitons~\cite{kevrekidis_discrete_2009}.  
To find them let's look for solution having the form $\alpha_{k} = \beta_k e^{i\omega z}$. Here $\beta_k$ are real numbers that doesn't depend on $z$. 
For $\alpha_{k}$ to satisfy~\eqref{eq:classic} these numbers should obey next equations:
\begin{equation}\label{eq:alg}
	-\omega \beta_k +\beta_{k+1 } + \beta_{k+1} + \beta^{3}_k = 0.
\end{equation}
For equations~\eqref{eq:alg} there is a number of solutions sets parameterized by parameter $\omega$  and which exist only when $\omega > 2$.
Some of these sets are linearly stable if $\omega$ is large enough and we will use them as boundary condition for~\eqref{eq:second_order}.  
On the top row of Fig.~\ref{fig:structure} some examples of such solutions are depicted.  
It should be noted that the equations~\eqref{eq:classic} can be considered a discrete analogue of the nonlinear Schrödinger equation~(NLSE). 
For this equation there is a well-known soliton solution quantum properties of which have been studied in several papers~\cite{corney_quantum_2001,lai_entangled_2009}. 
Discrete analog of such solution is shown on Fig.~\ref{fig:structure}(a). 
In case of infinite number of waveguides and $\omega\rightarrow 2$ mentioned discrete soliton should become similar to NLSE-soliton and have the same quantum properties. 
In the further discussion, we will focus on situations where discrete effects are significant. 
This means that we will use discrete solitons corresponding to large values of $\omega$ and having no continuous analogs, as shown in Fig.~\ref{fig:structure}(b) and Fig.~\ref{fig:structure}(c).
By solving equations~\eqref{eq:second_order} we want to study the quantum properties of the light field.
First of all we are interested in the entanglement that forms between the modes of different optical waveguides.
We will limit our consideration only by studying bipartite entanglement~\cite{gebremariam_tesfahannes_generation_2020, tura_detecting_2014}.
For this purpose we will use quantity called logarithmic negativity~($E_N$)~\cite{vidal_computable_2002} and will calculate it for different pairs of waveguides. 
In case of Gaussian quantum states $E_N$ for two waveguides with indexes $k$ and $l$ can be calculated based on correlation matrix $\bm{\sigma}$:
\begin{equation}\label{eq:cov_matr}
	\sigma_{m,n}=\langle\hat{\xi}_{m}\hat{\xi}_{n}+\hat{\xi}_{m}\hat{\xi}_{n}\rangle\bigl/2\bigr.-\langle\hat{\xi}_{m}\rangle\langle\hat{\xi}_{n}\rangle,    
\end{equation}
where indexes $m$ and $n$ take value from $1$ to $4$, and $\hat{\xi}_{m}$ and $\hat{\xi}_{n}$ are the corresponding components of a four-dimensional vector $\hat{\bm{\xi}}=\left(\,\hat{q}_{k},\hat{p}_{k},\,\hat{q}_{l},\,\hat{p}_{l}\,\right)^{T}$:
\begin{equation}\label{eq:qp}
	\hat{q}_{k,l}=\frac{1}{\sqrt{2}}\left(\hat{\psi}^{\dag v}_{k,l}+\hat{\psi}^{v}_{k,l}\right),\hspace{2.5mm}\hat{p}_{k,l}=\frac{i}{\sqrt{2}}\left(\hat{\psi}^{\dag v}_{k,l}-\hat{\psi}^{v}_{k,l}\right).
\end{equation}
\begin{figure}[h!]
	\centering\includegraphics[width=7cm]{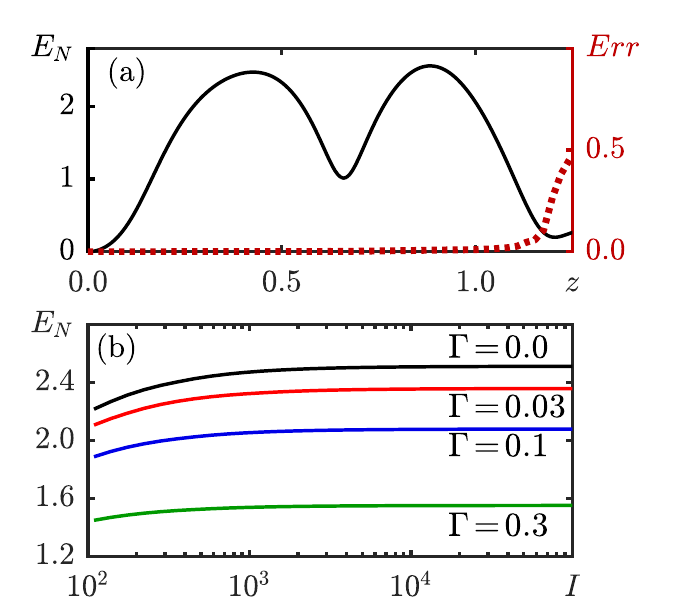}
	\caption{(a) Evolution of quantum properties of light propagating along an array of nonlinear waveguides. 
		The black solid line corresponds to the evolution of logarithmic negativity, the red dotted line is the evolution of the parameter that determines the applicability of the Gaussian approximation. (b) Dependence of the maximum value of $E_N$ attained during the propagation of light along the array on the light intensity in the central waveguide. All curves are calculated for a soliton of~Fig.\ref{fig:structure}(b) and $\omega=10.0$} 
	\label{fig:results}
\end{figure}
using matrix~\eqref{eq:cov_matr} it is possible to calculate logarithmic negativity:
\begin{equation}\label{eq:EN}
	E_N = -\dfrac{1}{2}\sum_r \rm{log}_2 \rm{min}\left\{1, 2\left|l_r\right|\right\},
\end{equation}
where $l_r$ are symplectic eigenvalues of matrix $\bm{\sigma}$.
\par
Before we discuss the results of calculations for quantum entanglement we should do a remark about Gaussian approximation we are using. 
The problem is that equations~\eqref{eq:heis} do not preserve the property of a quantum state to be Gaussian. 
But the initial quantum state is Gaussian. 
So we may assume that our approximation is valid for a certain distance of pulse propagation. 
And we need to estimate this distance to make sure the results we get are correct.
To solve this problem we consider the next order of cumulants expansion. 
We let $\ll\hat{X}_1\cdot\hat{X}_2\cdot\hat{X}_3\gg$ to be non zero but higher order cumulants still remains zero.
In this assumption we can derive the set of equations to calculate third order cumulant and compare it with zero.
We calculate next quantity:
\begin{equation}
	Err \!= \!\!\dfrac{\rm{max}\!\left\{\underset{k,l}{\rm{max}}\!\left|\ll\!\hat{\psi}^{\dag v}_{k}\!\cdot\!\hat{\psi}^{ v}_{k}\!\cdot\!\hat{\psi}^{ v}_{l}\!\!\gg\right|\!,\underset{k,l}{\rm{max}}\!\left|\ll\!\hat{\psi}^{ v}_{k}\!\cdot\!\hat{\psi}^{ v}_{k}\!\cdot\!\hat{\psi}^{ v}_{l}\!\!\gg\right|\right\}}
	{\rm{max}\left\{\underset{k,l}{\rm{max}}\left|\langle\hat{\psi}^{\dag v}_{k}\!\cdot\!\hat{\psi}^{ v}_{k}\!\cdot\!\hat{\psi}^{ v}_{l}\rangle\right|,\underset{k,l}{\rm{max}}\left|\langle\hat{\psi}^{ v}_{k}\!\cdot\!\hat{\psi}^{ v}_{k}\!\cdot\!\hat{\psi}^{ v}_{l}\rangle\right|\right\}}\!.
\end{equation}
Here we divide the maximum value between all cumulants we need to derive equations~\eqref{eq:second_order} by maximum value of averages which we calculate using the fact that corresponding cumulant is zero.
Thus, the Gaussian approximation is valid when the value of $Err$ is much less than one.
Dependency of $Err$ on the distance of light propagation along the array is shown on Fig.~\ref{fig:results}(a) by red dotted line. It is calculated for discrete soliton having filed distribution like the one shown on Fig.~\ref{fig:structure}(b) and parameters: $\omega = 10$, $L=0.01$.
From the curve one may notice that the defined quantity grows very slow for $z<1$, but after that point it increases dramatically.
This means that our approximation is valid until propagation distance is less than one.
This threshold distance depends weakly on the discrete soliton type and parameters of the system $\omega$ and $L$. 
All further calculations will be limited by this distance.
\par
Our calculations show that the distribution of the average photons number between waveguides does not change on propagation distance for which Gaussian approximation is valid. 
It remains the same as it is on the input of an array. 
At the same time the discussed distance is enough for the quantum state to change enough to form entangled states.
An example of the evolution of logarithmic negativity~($E_N$) is shown in Fig.~\ref{fig:results}(a) with a black solid line.
It is calculated for two central waveguides of a discrete soliton with an amplitude distribution as in Fig.~\ref{fig:structure}(b), and the parameters are $\omega=10.0$ and $L=0.01$.
The discussed curve shows that $ E_N $ reaches its maximum value rather quickly. The maxima point is characterized by the greatest degree of entanglement that can be achieved in the system. 
For a given propagation distance, we will discuss further results.
Distributions of $E_N$ calculated for different pairs of waveguides and three types of discrete solitons are shown on Fig.~\ref{fig:structure} on the bottom line.
These images show that entanglement is formed mainly between central waveguides having maximal intensity.
For discrete analog of NLSE-soliton~(Fig.~\ref{fig:structure}(a,d)) bipartite entanglement is present only between the central waveguide and its nearest neighbors.
Moreover, as $\omega$ increases, the distribution of the average number of photons degenerates to a situation where almost all photons are in one waveguide. 
Of course, in such a situation, there is no entanglement in the system.
For the case of sign changing discrete solitons like depicted on Fig.~\ref{fig:structure}(b,c) bipartite entanglement present only for the pair of optical modes having maximal average photon number~(see Fig.~\ref{fig:structure}(e,f)). 
In our opinion, solitons of this kind are the most interesting from a practical point of view.
Our calculations show that when the complex amplitude of a soliton changes sign several times~(as in Fig.~\ref{fig:structure}(c)), entanglement is formed only between distinct pairs of waveguides.
In practice this fact can be used to construct optical integrated circuit generating multiple entangled states in parallel.
\par
For most optical materials, the cubic nonlinearity coefficient is relatively small. 
So, to form a discrete soliton, it is necessary to use laser pulses with a large number of photons.
It is important to understand what happens to the discussed entangled states when the intensity of the incoming light increases and we move to the limit that is usually considered classical.
To achieve this goal we should perform calculations for $L$ tending to zero.
The result is shown in Fig.~\ref{fig:results}(b) by black line. 
An interesting fact here is that the quantum state remains entangled as the intensity of the incoming light increases.
This result can be easily explained.
If we look at equations~\eqref{eq:second_order} we can notice that  variables $\Delta^{n}_{k,l}$ and $\Delta^{a}_{k,l}$ are small for small values of $L$.
We can linearize them and get a system of inhomogeneous equations.
It is clear that solution is proportional to source term $\Delta^{n,a}_{k,l}\sim L \cdot \alpha^2_k$. 
At the same time, the correlation matrix~\eqref{eq:cov_matr} which determines entanglement, is proportional to $\bm{\sigma}\sim \Delta^{n,a}_{k,l}/L$.
In combination these two facts lead to that $\bm{\sigma}$ does not depend on $L$. 
From the point of view of quasiprobability distributions, this means that the evolution of quadratures does not depend on the position of the distribution center.
\par
The last thing we want to mention here is the influence of absorption on the process of entangled states formations.
To consider absorption, we use a standard model of interaction with   a reservoir of an infinite number of harmonic oscillators~\cite{drummond_quantum_2014}.
This will result in additional terms in ~\eqref{eq:second_order} which have the form $-\Gamma \alpha_k / 2$, $-\Gamma \Delta^{n}_{k,l}$ and $-\Gamma \Delta^{a}_{k,l}$ for corresponding equations sets. 
Here $\Gamma$ is the absorption coefficient.
In the presence of absorption, the evolution of entanglement remains qualitatively the same.
The only thing that changes is the maximum value of logarithmic negativity that can be achieved in the system.
On Fig.~\ref{fig:results}(b) dependency of this quantity on intensity of input beams is shown for several values of $\Gamma$.
The entangled states of light are still formed even for relatively high absorption coefficients. 
In real experiment setups coupling coefficient $\kappa/v$ can be about  $1 \rm{cm}^{-1}$~\cite{solntsev_generation_2014}.
For such value $\Gamma=0.3$ corresponds to absorption $0.3 \rm{cm}^{-1}$ which is very high for typical optical materials.
\par
In conclusion we underline the achieved results.
We have studied the quantum properties of discrete solitons propagating through an array of cubic nonlinear waveguides.
Certain types of such field distributions form entangled states of light between optical modes of multiple pairs of coupled waveguides.
We propose that they can be used to create integrated optical devices for generating continuous-variable entangled states.
The advantage of such scheme is that discussed quasi-solitons are stable and do not require high quality of light field amplitude matching on the input of the array.
Also we have shown that entangled states are formed even in the case of using input sources with high intensities.
This means that proposed scheme does not require any specific materials with extremely high nonlinear coefficients.
Absorption which is present in all real materials also does not influence the entangled states generation process too much. 
There is an open question of the influence of the other sources of noise on the discussed process, such as Raman scattering or fluctuations of coupling coefficient.
However, this is the topic for future studies.
\section*{Acknowledgments}
The reported study was funded by RFBR according to the research project № 19-32-80038
\bibliographystyle{apsrev4-1}
\bibliography{DiscreteSolitons}
\end{document}